\documentclass[twocolumn]{aastex631}

\usepackage{amssymb}
\usepackage{amsmath}
\usepackage{booktabs}
\graphicspath{{figures/}}
\usepackage{subfigure}

\newcommand{\brunt}{Brunt-V\"ais\"al\"a }
\newcommand{\mearth}{M$_{\oplus}$}
\newcommand{\Zsolar}{Z$_{\odot}$}

\usepackage{multirow}
\usepackage{soul}

\begin{document}

\title{The Evolution of Jupiter and Saturn as a function of the R$_{\rho}$ Parameter}

\correspondingauthor{Ankan Sur}
\email{ankan.sur@princeton.edu}

\shorttitle{Giant Planets from Schwarzschild to Ledoux}
\shortauthors{Sur et al.}

\author[0000-0001-6635-5080]{Ankan Sur}
\author[0000-0002-3099-5024]{Adam Burrows}
\author[0000-0001-6708-3427]{Roberto Tejada Arevalo}
\author[0000-0001-8283-3425]{Yubo Su}
\affiliation{Department of Astrophysical Sciences, Princeton University, 4 Ivy Lane,
Princeton, NJ 08544, USA}

\begin{abstract}
Computed using the \texttt{APPLE} planetary evolution code, we present updated evolutionary models for Jupiter and Saturn that incorporate helium rain, non-adiabatic thermal structures, and ``fuzzy" extended heavy-element cores. Building on our previous Ledoux-stable models, we implement improved atmospheric boundary conditions that account for composition-dependent effective temperatures and systematically explore the impact of varying the parameter $R_{\rho}$, which allows one to explore in {an approximate way the efficiency of semiconvection}. For both Jupiter and Saturn, we construct models spanning from $R_{\rho}=1$ (Ledoux) to $R_{\rho}=0$ (Schwarzschild), and identify best-fit solutions that match each planet’s effective temperature, equatorial radius, lower-order gravitational moments, and atmospheric composition at 4.56 Gyr. We find that lower $R_{\rho}$ values lead to stronger convective mixing, resulting in higher surface metallicities and lower deep interior temperatures, while requiring reduced heavy-element masses and lower initial entropies to stabilize the dilute inner cores. Our Saturn models also broadly agree with the observed \brunt frequency profile inferred from Cassini ring seismology, with stable layers arising from both the helium rain region and the dilute core. These findings support the presence of complex, compositionally stratified interiors in both gas giants.
\end{abstract}
    
\section{Introduction}

Accurate delineation of convective boundaries is critical to any planetary evolution calculation, as in stellar astrophysics \citep{Chabrier1997, Anders2022}. Traditional models have uniformly assumed an adiabatic interior --- i.e., a compositionally homogeneous envelope in which {convection spans the entire planet \citep{Hubbard1968, Hubbard1969, Hubbard1970, Burrows1997, Baraffe1998, Burrows2001, Baraffe2003,marley2021_sonora}}. However, data returned by the Juno \citep{Bolton2017a} and Cassini \citep{Spilker2019} missions now point to stable heavy‐element gradients in the deep interiors of Jupiter and Saturn \citep{Wahl2017, Debras2019, Militzer2022, Howard2023, Militzer2024, Fuller2014, Mankovich2021}. Although currently unexplained, such stratification may originate from dilution and partial erosion of an initially compact core \citep{Guillot2004, Moll2017, Fuentes2023} or from direct head-on collisions of planetesimals onto planetary embryos \citep{Liu2019} (however, this formation mechanism has recently been challenged \citep{Meier2025}). In addition, helium phase separation (``helium rain", \citealp{Stevenson1977a}), first proposed to explain Saturn’s anomalous luminosity, can likewise produce a stably stratified region that impedes heat transport {through it \citep{Fortney2003, Pustow2016}}.

Rigorous treatment of convective transport is essential for interpreting the structural and atmospheric properties of both solar‐system giants and exoplanets, since convection governs heat redistribution and the vertical mixing of heavy elements, which in turn influences the observed planetary radius and atmospheric composition \citep{Basu2012, Bloot2023, Polman2024, Knierim2025}. Recent studies demonstrate that steep internal composition gradients {associated with lower interior entropies} can suppress large‐scale convective motions, thereby preserving an initially stably stratified core over Gyr timescales \citep{Knierim2024,TejadaArevalo2024b,Sur2025}, {and helping maintain a warmer internal temperature profile.} In the case of Jupiter and Saturn, self‐consistent evolutionary models by \cite{TejadaArevalo2024b} and \cite{Sur2025} that begin with a diffuse, ``fuzzy" core reproduce multiple observables --- such as current luminosity, lower order gravitational moments, atmospheric compositions, and interior density profiles — more accurately than traditional adiabatic models. Gradients in the composition of a planet affect the efficiency of heat transport throughout its interior, resulting in changes in its cooling history and mass–radius relationship \citep{Chabrier2007, Leconte2012, Leconte2013}. Moreover, the gradual mixing of heavy-element-rich regions into the overlying envelope controls the enrichment of volatiles in the atmosphere and places stringent constraints on the planet’s total heavy‐element inventory. Consequently, any robust planetary evolution framework must account for these inhibited convective processes and their impact on the long‐term redistribution of heavy elements.

Convective stability in planetary interiors is assessed using the Schwarzschild criterion \citep{Schwarzschild1958} when the composition is uniform, and the Ledoux criterion \citep{Ledoux1947} when there are non‐negligible composition gradients. In terms of temperature gradients, these criteria can be written as

\begin{align}
    &\nabla_T - \nabla_{\rm ad} < 0 \,\,\,(\rm Schwarzschild) \\ &\nabla_T - \nabla_{\rm ad} - \frac{\varphi}{\delta}\nabla_{\mu}< 0 \,\,(\rm Ledoux)\, ,
    \label{eq:schw_criterion1}
\end{align}
where $\nabla_T = d\ln T / d\ln P$, $\nabla_{\rm ad}$ is the adiabatic gradient, $\nabla_\mu = d\ln\mu / d\ln P$ is the mean‐molecular‐weight gradient, $\delta = -(\partial\ln\rho/\partial\ln T)_{P,\mu}$, and $\varphi = (\partial\ln\rho/\partial\ln\mu)_{P,T}$. Equivalently, one can express these stability conditions in terms of specific entropy gradients \citep[see][]{Tejada2024, Sur2024_apple}. 

{A layer is termed semiconvective if it is unstable according to the Schwarzschild condition (i.e., $\nabla_T > \nabla_{\rm ad}$), but remains stable according to the Ledoux criterion (i.e., $\nabla_T < \nabla_{\rm ad} + (\varphi/\delta),\nabla_\mu$). Numerous hydrodynamical studies have investigated the nature of semiconvection and the resulting mixing processes \citep{Rosenblum2011, Mirouh2012, Wood2013, Garaud2015, Moore2016, Moll2016, Moll2017, Fuentes2022, Fuentes2024, Fuentes2025}. However, despite these efforts, there is still no fully self-consistent implementation of semiconvection in one-dimensional stellar or planetary evolution codes.} 

{Typically, to evaluate the mode of energy transport, one introduces the density‐ratio parameter\footnote{{There is an ambiguity in the literature regarding the definition of $R_0$. Some authors define it as the temperature gradient relative to the composition gradient \citep[e.g.][]{Leconte2012, Moore2016, Moll2016}, whereas others take the inverse, i.e., composition over temperature \citep[e.g.][]{Spruit2013, Fuentes2024}. We don't use this quantity, but use a global parameterization that brackets the potential behavior due to semiconvection.}} \citep{Walin1964, Kato1966, Rosenblum2011, Mirouh2012, Wood2013, Leconte2012, Leconte2013, Garaud2015, Moll2016, Garaud2018, Fuentes2022, Fuentes2025}}

\begin{equation}
    R_0 = \frac{\delta}{\varphi}\frac{\nabla -\nabla_{\rm ad}}{\nabla_\mu} ,
    \label{eq:R0}
\end{equation}
{which measures the ratio of the destabilizing thermal stratification to the stabilizing compositional stratification. The quantity $R_0$ is a structural diagnostic that can be computed locally for every fluid parcel and determines whether the region is stable, convective, or semiconvective. In this work, however, we adopt an alternative parameterization that does not explicitly model semiconvective energy transport. Instead, following previous studies \citep{Mankovich2016, Mankovich2020}, we introduce a global parameter $R_{\rho}$ to modify the convective stability criterion as}:

\begin{equation}
    \nabla - \nabla_{\rm ad} - R_{\rho}\frac{\varphi}{\delta}\nabla_{\mu} < 0.
    \label{eq:R_rho}
\end{equation}

{The parameter $R_{\rho}$ is not equivalent to the density ratio $R_0$, but it brackets the two possible extremes: $R_{\rho}=0$ recovers the Schwarzschild criterion, in which any semiconvective regions become fully convective in short timescales, whereas $R_{\rho}=1$ corresponds to the Ledoux criterion, for which there is no semiconvection\footnote{{Note that by combining Equations \ref{eq:R0} and \ref{eq:R_rho}, one finds that convective instability occurs when $R_0 - R_\rho > 0$. Thus, a larger $R_0$ or a smaller $R_\rho$ both signify more efficient energy transport. This distinction is often obscured by the similar notation, even though $R_0$ and $R_\rho$ represent fundamentally different physical quantities.}}.
This parameterized $R_{\rho}$-based approach has been adopted in previous studies \citep{Mankovich2016, Mankovich2020, Howard2024, Sur2024_apple, Nettelmann2025}, and we extend this established framework to systematically explore the evolution of Jupiter and Saturn with fuzzy cores in this context. We emphasize, however, that we are not claiming to address here semiconvection, per se, given the remaining significant uncertainties in the treatment of doubly-diffusive instabilities, merely to be bracketing the evolutionary and structural possibilities.}

{While the $R_{\rho}$ prescription simplifies the physics by assuming only two modes of energy transport — fully convective or stable — it retains a clear physical interpretation. If semiconvective transport is inefficient, the convective stability criterion approaches the Ledoux limit (\( R_{\rho} \to 1 \)); if it is efficient, it mimics Schwarzschild instability (\( R_{\rho} \to 0 \)) by transporting energy and composition effectively. Although \( R_0 \) (a structural diagnostic) and \( R_{\rho} \) (a modeling parameter) are distinct, they are related: a layer is treated as convective if \( R_0 > R_{\rho} \), and stable otherwise. Thus, \( R_{\rho} \) sets the threshold for convective onset in the presence of a stabilizing composition gradient. In models that treat semiconvection as a distinct transport process, the critical density ratio \( R_{\mathrm{crit}} \) at which semiconvection sets in plays a similar role: lower \( R_{\mathrm{crit}} \) or \( R_{\rho} \) leads to more mixing. However, \( R_{\rho} \) is an effective tuning parameter, while \( R_{\mathrm{crit}} \) is a physical threshold measurable in laboratory experiments, and their values need not match.}

In the past, the {density ratio $R_0$ has at times been used} to model semiconvection in giant planets. \cite{Nettelmann2015} investigated Jupiter’s thermal evolution by including double‐diffusive convection and helium rain, {in what remains the only study to self-consistently relate the heat flux to the superadiabaticity prescription of \citet{Wood2013}}. By adjusting the hydrogen–helium phase diagram of \citet[][LHR0911]{Lorenzen2009, Lorenzen2011}, they explored shifts in the miscibility curve and determined the total heat flux as a function of superadiabaticity and semiconvective layer height, to match Jupiter’s observed effective temperature. {On the other hand, \citet{Mankovich2016, Mankovich2020} modeled the evolution of both Jupiter and Saturn using a parameterized approach in which $R_{\rho}$ was treated as a tunable quantity controlling convective efficiency. They found that Jupiter’s interior is superadiabatic (with $R_{\rho}=0.05$) throughout the helium‐gradient layer, while Saturn could exhibit regions that are either adiabatic or superadiabatic.} A comparable conclusion was reached by \cite{Howard2024}, who constrained $R_{\rho} \leq0.1$. However, these previous studies considered only the effect of superadiabaticity in helium‐rain regions; they did not include any heavy‐element gradients or a diluted core in their models. In addition, {except for \citet{Howard2024}, who used the \cite{Chabrier2019} equation of state with \cite{Howard2023} correction, most} relied on older equations of state and atmospheric boundary conditions rather than the latest available data. Furthermore, their models make no effort to engage with the constraints provided by Juno or Cassini observations.

Convective regions can drive mixing into adjacent, stably stratified layers through convective boundary mixing \citep{Korre2019, Anders2023}. When only composition is mixed beyond the boundary, the process is called ``convective overshoot." If both entropy and composition are mixed, it is termed ``convective penetration.'' \citet{Anders2022}, using 3D hydrodynamical simulations in a stellar context, showed that over long timescales, the Schwarzschild and Ledoux criteria yield the same convective boundaries. Their results suggest that convective overshoot can erode nearby semiconvective layers as buoyant plumes entrain less dense material, causing the convective region to expand. This entrainment continues until the two stability criteria converge. This mechanism would be especially relevant for Jupiter and Saturn, whose interiors are thought to contain stably stratified regions \citep{Fuller2014, Mankovich2019, Fortney2023}. Additionally, \citet{Tulekeyev2024} found that convective cells tend to grow and merge, potentially making the entire planet convective. This challenges the persistence of fuzzy cores, particularly in Saturn. {However, the rapid rotation of gas giants can reduce the kinetic energy flux available for compositional mixing, significantly extending the mixing timescale of semiconvective layers \citep{Fuentes2024, Fuentes2025}. The latter studies demonstrate, for the first time, semiconvection in a full-sphere geometry, with and without rotation, across different density ratios.}

In the present study, we adopt the implementation described above and in \cite{Sur2024_apple}, in which the parameter $R_{\rho}$ governs the transition between convective and non-convective regions. By varying $R_{\rho}$, we examine how different convective stability prescriptions (and the semiconvective processes they model) affect the long‐term cooling, luminosity evolution, and heavy‐element redistribution in Jupiter and Saturn. This paper is organized as follows: Section~\ref{sec:method} provides a brief overview of our numerical code and the modules employed. Section~\ref{sec:models} presents the evolutionary models of Jupiter and Saturn under different convection criteria and explores the effects of varying $R_{\rho}$. Finally, Section~\ref{sec:conclusions} summarizes our key findings and offers broader discussion and interpretation.

\begin{figure*}
    \hspace{-0.4cm}\includegraphics[width=1.05\linewidth]{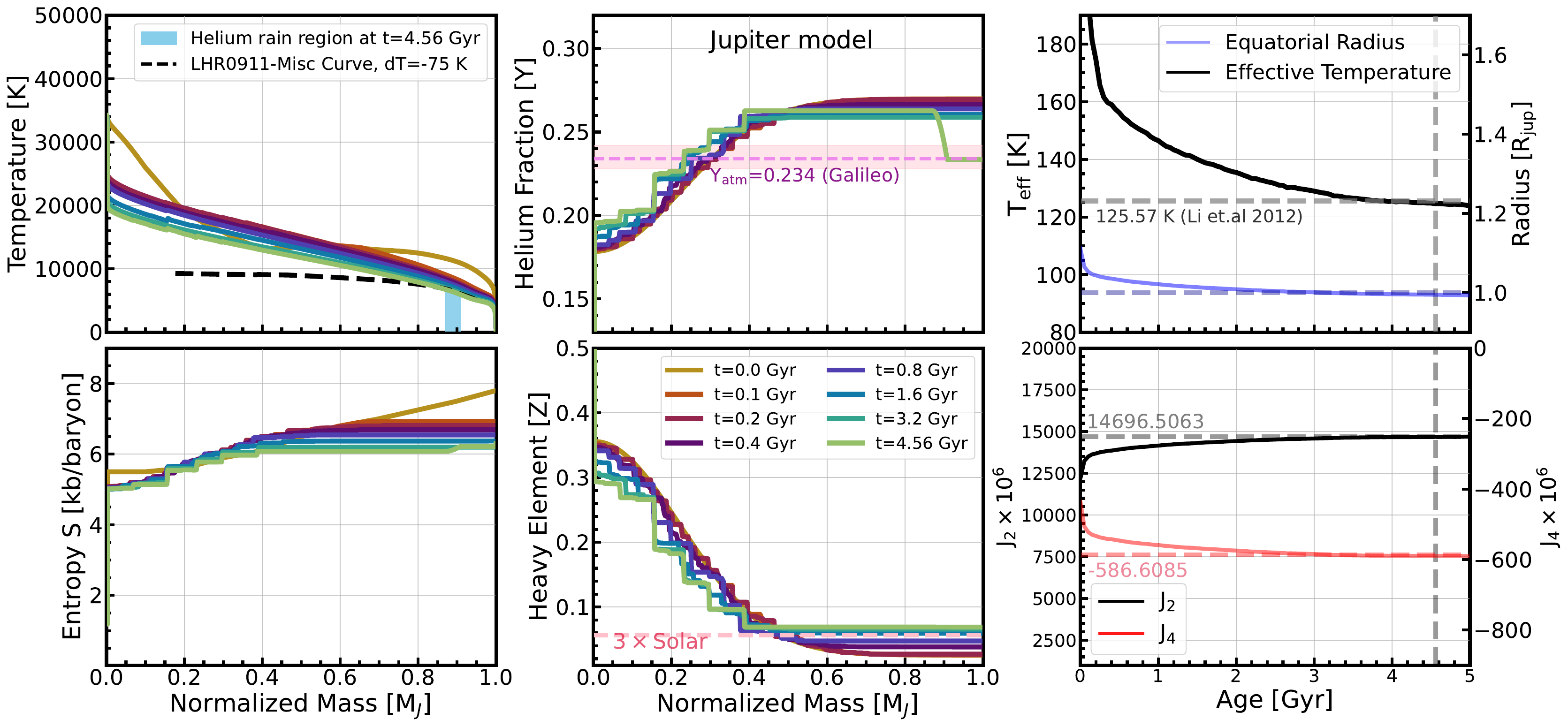}
    \caption{Evolutionary model for Jupiter with an initial fuzzy core that experiences Schwarzschild convection ($R_{\rho}=0$). This model reproduces, within {$\sim0.7$\% of the various measured quantities}, the current values of the effective temperature, equatorial radius, atmospheric helium abundance, outer envelope metallicity, and $J_2$ and $J_4$ gravitational moments. The initial outer entropy is 7.8 k$_{\rm B}$/baryon, while the interior entropy is 5.8 k$_{\rm B}$/baryon. The panels display: (top left) the evolution of the temperature profile; (top middle) the evolution of the helium mass fraction ($Y$) profile; (top right) the evolution of the effective temperature and equatorial radius; (bottom left) the evolution of the entropy profile; (bottom middle) the evolution of the $Z$ profile; and (bottom right) the evolution of the gravitational moments $J_2$ and $J_4$. This Jupiter model contains 38 \mearth of heavy elements, including a compact core of 1 \mearth ($Z = 1$). The outer metallicity starts at 1.9 \Zsolar\ and increases to 3.66 \Zsolar. Helium rain begins at $\sim 3.5$ Gyrs, based on the LHR0911 miscibility curves with a -75 K temperature shift, resulting in outer helium depletion to $Y = 0.233$, consistent with \textit{\textit{Galileo}} entry probe measurements \citep{vonZahn1998}.}
    \label{fig:jupiter_schw_best_fit}
\end{figure*}

\section{Method} 
\label{sec:method}
We use \texttt{APPLE} \citep{Sur2024_apple}, which solves the hydrostatic structure with energy and species transport implicitly in time. For both planets, we adopt the most recent hydrogen-helium equation of state from \cite{Chabrier2021}, and heavy elements in the envelope are modeled using the AQUA equation of state \citep{Haldemann2020}. Our dilute core of mass $M_z$ consists of a small, compact core (of mass $M_c$) embedded within an extended heavy‐element distribution in the surrounding envelope \footnote{In our models, $M_z$ denotes the total heavy-element mass in the planet, which includes a central compact core of mass $M_c$. The heavy-element mass in the envelope is therefore $M_z^{\rm env} = M_z - M_c$.}. The compact core itself is composed of iron and post‐perovskite \citep{Keane1954, Stacey2004, Zhang2022}. 

Helium rain is modeled by adopting the hydrogen–helium miscibility curves of \citet[LHR0911]{Lorenzen2009, Lorenzen2011}, with a user‐specified temperature shift to explore the sensitivity of phase separation. At each timestep, we compare the local temperature–pressure profile to the shifted demixing curve; whenever the envelope becomes supersaturated in helium, we compute the resulting fractionation and use our diffusion algorithm from \cite{Sur2024_apple} to rain out helium. 

For the outer boundary condition, we employ the updated atmospheric boundary tables of \citet{Chen2023}. Unlike previous implementations (e.g., \citealt{Sur2024_apple, Sur2025}), which assumed a fixed helium abundance in the atmosphere, our current calculation interpolates the boundary conditions over both helium and heavy‐element mass fractions, thereby capturing variations in mean molecular weight and opacity.

By solving the hydrostatic structure and heat and composition transport simultaneously—in conjunction with the chosen core composition, helium‐rain prescription, and atmosphere boundary tables—\texttt{APPLE} computes self‐consistent evolutionary sequences for Jupiter and Saturn, tracking radius, luminosity, rotation rate, and internal composition profiles over time. 

For Jupiter, we explored a wide parameter space by varying the total heavy-element mass between 38 and 45 \mearth, the compact core mass between 1 and 5 \mearth, the temperature shift applied to the LHR0911 hydrogen-helium miscibility curve from –200 K to +500 K, and the initial surface entropy between 7.5 and 8.5 k$_\mathrm{B}$/baryon. Rather than prescribing a fixed functional form, the internal entropy profiles were manually constructed to ensure thermodynamic consistency across the planet’s structure. For each value of $R_{\rho}$ considered in this study, we adjusted the initial entropy profile accordingly and identified the best-fit parameters that reproduce Jupiter’s key observables. We applied the same modeling framework to Saturn to investigate how varying $R_{\rho}$ and {the initial thermal stratification influences its long-term thermal and compositional evolution}. The total heavy-element content was varied between 23 and 27 \mearth, with a compact core mass ranging from 1 to 5 \mearth, and initial surface entropies spanning 7.0 to 8.5 k$_\mathrm{B}$/baryon. We adopted the fixed temperature shifts in the LHR0911 miscibility curve for Saturn to remain consistent with the Jupiter models, rather than independently tuning them. Overall, we ran over 4,000 simulations for each planet to thoroughly sample the parameter space and identify robust evolutionary solutions.

\begin{figure*}
    \hspace{-0.4cm}\includegraphics[width=1.05\linewidth]{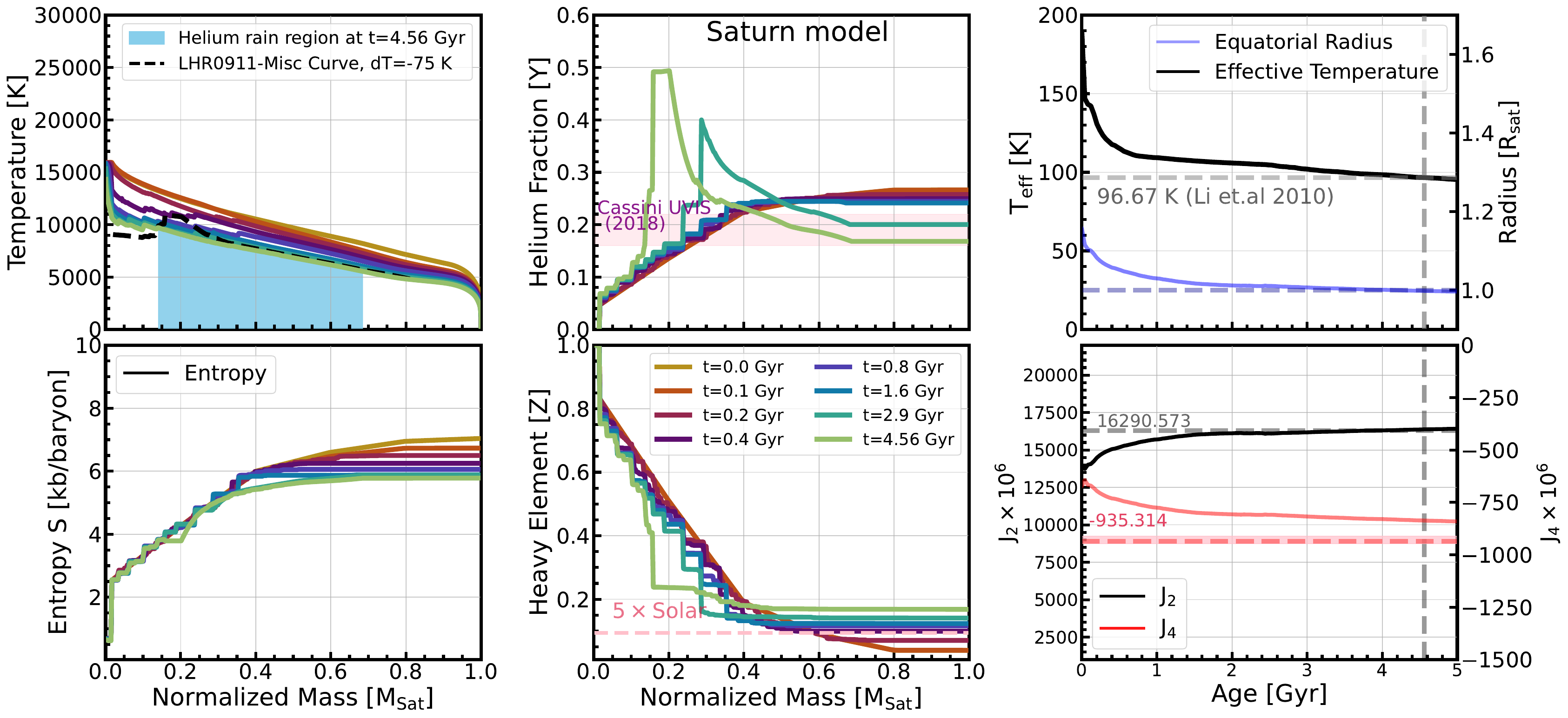}
    \caption{Best-fit Saturn evolutionary model assuming an initially fuzzy core and Schwarzschild convection ($R_{\rho}=0$). This model reproduces Saturn’s present-day effective temperature, equatorial radius, outer envelope metallicity, and gravitational moment, $J_2$ {within 0.5\% of their measured values}. The initial entropy is 7.0 k$_\mathrm{B}$/baryon at the outer boundary and $\sim$4.0 k$_\mathrm{B}$/baryon in the deep interior. The six panels mirror those shown in Figure~\ref{fig:jupiter_schw_best_fit}: (top left) temperature profile evolution; (top middle) helium mass fraction ($Y$) profile; (top right) evolution of effective temperature and equatorial radius; (bottom left) entropy profile; (bottom middle) heavy-element mass fraction ($Z$) profile; and (bottom right) gravitational moments $J_2$ and $J_4$. The model contains a total of 24.5 $M_\oplus$ in heavy elements, including a compact 1.5 $M_\oplus$ core. The outer envelope metallicity rises from an initial value of 2.0 $Z_\odot$, to 9 $Z_\odot$ due to interior mixing. The LHR0911 hydrogen–helium miscibility curve shifted by -75 K (equivalent to the corresponding Jupiter model) leads to an atmospheric helium mass fraction of $Y_{\rm atm} \sim 0.168$, in agreement with constraints from \citet{Conrath2000} and \citet{Koskinen2018} (pink shaded region in the top middle panel). The miscibility boundary and helium rain region for the current epoch of the best-fit model are highlighted using a blue field in the top left panel.}
    \label{fig:saturn_best_fit}
\end{figure*}  

\section{RESULTS}
\label{sec:models}

Our earlier work \citep{Sur2025} demonstrated the importance of incorporating fuzzy cores via the Ledoux stability criterion, showing that a combination of helium rain and compositional gradients can account for the observed properties of Jupiter and Saturn. In this study, we build on that framework by systematically varying $R_{\rho}$, comparing evolution from the Ledoux to Schwarzschild convection criterion and quantifying how the degree of semiconvective mixing (crudely parameterized by $R_\rho$) impacts the thermal and structural evolution, as well as the required initial thermal profiles.

\subsection{Jupiter Model}
\label{jupiter}

We first revisit Jupiter’s thermal history using updated atmospheric boundary conditions that allow the effective temperature to depend on the helium and heavy element mass fractions at the surface \citep[following][]{Chen2023}. Incorporating these refinements into our models, we recalculated the evolution for different values of $R_{\rho}$. 

Our best-fit Schwarzschild-stable Jupiter model contains 38 \mearth\ of heavy elements, with a compact core of 1 \mearth. This configuration successfully reproduces Jupiter’s equatorial and mean radii. A temperature shift of –75 K in the LHR0911 hydrogen-helium miscibility curve enables the model to match the Galileo probe’s measurement of an atmospheric helium mass fraction of $Y_{\rm atm} = 0.234$ \citep{vonZahn1998}. To reproduce Jupiter’s effective temperature of $\sim$125 K at 4.56 Gyr, the model requires a surface entropy of $S_{\rm atm} \sim 7.8$ k$_\mathrm{B}$/baryon and a deep interior entropy of $S_{\rm int} \sim 5.8$ k$_\mathrm{B}$/baryon, implying initial conditions that are colder than our earlier Ledoux-based solutions. The planet’s central temperature begins at approximately 35,000 K and cools to around 20,000 K by the present epoch. The final surface metallicity reaches 3.6 \Zsolar, consistent with {observational constraints \citep{Guillot2023}}. The evolutionary history is shown in Figure \ref{fig:jupiter_schw_best_fit}.

For comparison, the best-fit Ledoux-stable model has a slightly higher heavy-element mass of 44.25 \mearth, including a 5 \mearth\ compact core. Due to the reduced efficiency of convective heat transport, this model requires a higher surface entropy of $\sim8.12$ k$_\mathrm{B}$/baryon and a +300 K shift in the LHR0911 miscibility curve to reproduce the observed helium depletion. Despite achieving similar present-day observables, the Ledoux and Schwarzschild models differ notably in their current and initial thermal profiles, internal entropy profiles, and inferred solid-core properties. 
\begin{figure*}
    \hspace{-0.4cm}\includegraphics[width=1.04\linewidth]{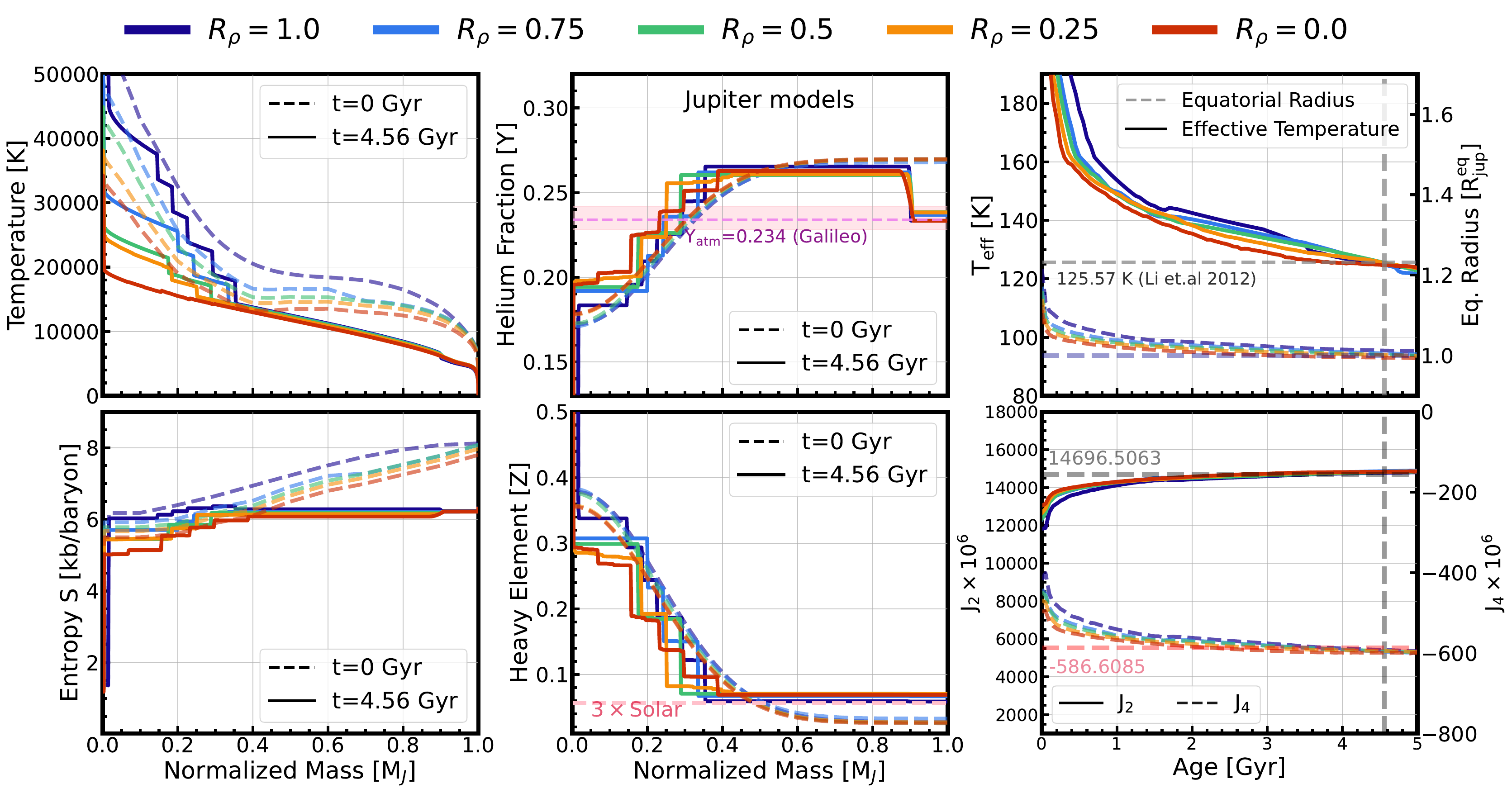}
    \caption{Best fit evolutionary models for Jupiter with varying $R_{\rho}$. Only the initial and final (4.56 Gyr.) timestamps are shown for comparison. The dark blue curve corresponds to our updated Ledoux model ($R_{\rho} = 1$), while the red curve shows the Schwarzschild model ($R_{\rho} = 0$). The light blue, green, and orange curves represent intermediate cases with $R_{\rho} = 0.75$, $0.50$, and $0.25$.  All quantities are within {1$\sigma$ of their measured central values.}}
    \label{fig:jupiter_rho_comp}
\end{figure*}

\begin{figure*}
    \hspace{-0.4cm}\includegraphics[width=1.04\linewidth]{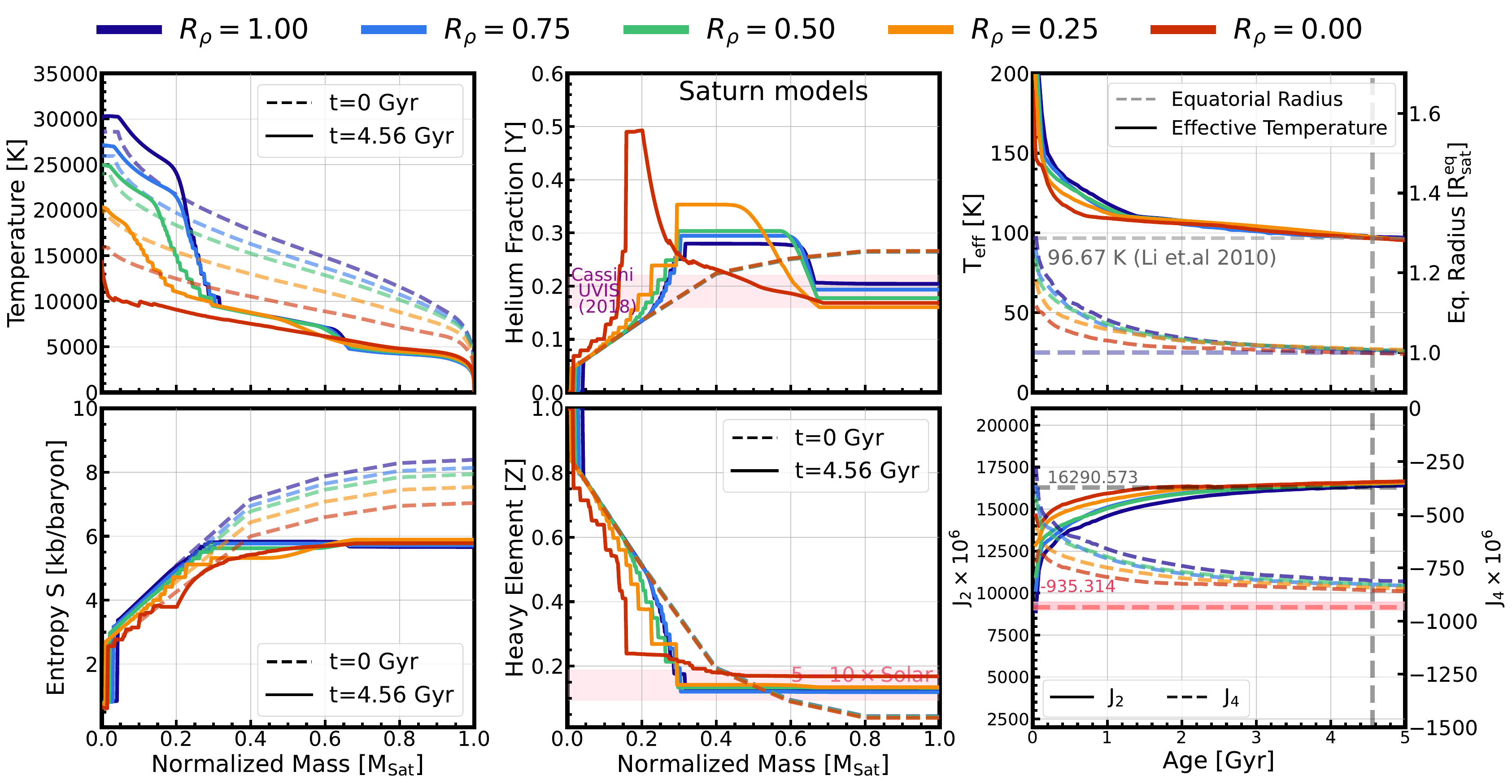}
    \caption{Same as Figure~\ref{fig:jupiter_rho_comp}, but for Saturn: best-fit evolutionary models with varying $R_{\rho}$ values. As $R_{\rho}$ decreases, the models exhibit higher effective temperatures during early evolution, enhanced final envelope metallicities, broader helium rain regions, and more helium depletion at the surface. The total heavy-element mass and initial entropies are adjusted accordingly to reproduce Saturn’s present-day structure, with the Schwarzschild model requiring lower initial central temperatures and a more compact core.}
    \label{fig:saturn_rho_comp}
\end{figure*} 
 
\subsection{Saturn Model}
\label{saturn}

The best-fit Schwarzschild-stable model for Saturn requires a total of 24.5 \mearth\ of heavy elements, including a 1.5 \mearth\ compact rocky core. Stronger mixing relative to the Ledoux case reduces the extent of the dilute core, leading to more efficient erosion of the compositional gradient and an increase in the final envelope metallicity to approximately 9 \Zsolar. The imposed miscibility shift yields a final atmospheric helium abundance of 0.168, broadly consistent with Cassini UVIS constraints \citep{Koskinen2018}. The region undergoing helium rain spans from 1 Mbar to about 7 Mbar, with a substantial amount of helium settling near the base of the stably stratified region. To match Saturn’s observed luminosity, the model requires an initial surface entropy of $\sim$7.0 k$_\mathrm{B}$/baryon and an interior entropy of $\sim$4.0 k$_\mathrm{B}$/baryon. The central temperature begins at $\sim$16,000 K and cools modestly to 14,000 K by 4.56 Gyr, indicating limited core cooling over solar-system timescales.  The evolutionary history is shown in Figure \ref{fig:saturn_best_fit}.

In contrast, our Ledoux-stable model for Saturn manifests a slightly higher total heavy-element content of 25.5 \mearth, with a diffuse core extending to roughly 50\% of Saturn’s radius. The compact core in this model is larger (4 \mearth), and the atmospheric helium abundance reached a slightly higher value of $\sim$0.204.{This is consistent with recent formation and evolution models that include helium rain, which predict Saturn’s atmospheric helium abundance to be around $\sim$0.2 and the heavy element mass in the envelope to be 23.2 \mearth\citep{Bodenheimer2025}}. Due to suppressed convective mixing, helium rain in the Ledoux model is more localized, and interior cooling is further inhibited. The initial surface entropy required to match Saturn’s present effective temperature is 8.35 k$_\mathrm{B}$/baryon, with a central temperature of 30,000 K at early times—hotter than in the Schwarzschild case. These differences highlight the sensitivity of Saturn’s structure and thermal state to the assumed convective regime.

\begingroup
  \setlength{\tabcolsep}{2pt}
  \renewcommand{\arraystretch}{1.2}
\begin{table}
\centering
\caption{Model input parameters for Jupiter and Saturn for different $R_{\rho}$ values. For each planet, we list the total heavy-element mass ($M_z$), compact core mass ($M_c$), temperature shift applied to the LHR0911 H/He miscibility curve ($\Delta T$), and the {initial outer entropy ($S^{\rm ini}_{\rm atm}$ in $k_{\rm B}$/baryon).}}
\label{tab:input_parameters}
\begin{tabular}{@{}lcccccc@{}}
\hline
\hline
Planet & Parameter & $R_{\rho}=0$ & $0.25$ & $0.5$ & $0.75$ & $1.0$ \\
\hline
\multirow{4}{*}{Jupiter}
  & $M_z$ [M$_\oplus$]       & 38.0 & 38.5 & 40.0 & 42.0 & 44.25 \\
  & $M_c$ [M$_\oplus$]       & 1.0  & 1.0  & 1.0  & 1.0  & 5.0 \\
  & $\Delta T$ [K] (LHR0911)   & –75  & +50  & +150 & +250 & +300 \\
  & $S^{\rm ini}_{\rm atm}$ [$k_{\rm B}$/baryon] & 7.80  & 7.96  & 8.08  & 8.08  & 8.12 \\
\hline
\multirow{4}{*}{Saturn}
  & $M_z$ [M$_\oplus$]       & 24.5 & 24.5 & 25.0 & 25.0 & 25.5 \\
  & $M_c$ [M$_\oplus$]       & 1.5  &  1.0  & 2.0  & 3.0  & 4.0 \\
  & $\Delta T$ [K] (LHR0911)   & –75  & +50  & +150 & +250 & +300 \\
  & $S^{\rm ini}_{\rm atm}$ [$k_{\rm B}$/baryon] & 7.0  & 7.5  & 7.9  & 8.1  & 8.35 \\
\hline
\end{tabular}
\end{table}
\endgroup

\subsection{Variation with $R_{\rho}$}
\label{sec:variation R}

To quantify how the degree of semiconvective mixing affects the thermal and compositional structure of Jupiter and Saturn, we systematically vary the parameter $R_{\rho}$ from 1.0 (Ledoux criterion) to 0.0 (Schwarzschild criterion). In addition to these two end-member cases, we explore intermediate regimes with $R_{\rho} \in \{0.75, 0.5, 0.25\}$ to capture the continuum of semiconvection efficiencies. For each value of $R_{\rho}$, we construct self-consistent evolutionary models by adjusting the total heavy-element mass, core mass, initial entropy profile, and the temperature shift applied to the LHR0911 hydrogen-helium miscibility curve. A full summary of the input parameters adopted for each model is provided in Table \ref{tab:input_parameters}, with corresponding model outcomes detailed in Table \ref{tab:jup_sat_comparison}. Figures \ref{fig:jupiter_rho_comp} and \ref{fig:saturn_rho_comp} show the initial interior structures and the evolutionary outcomes at 4.56 Gyr for Jupiter and Saturn, respectively, as a function of $R_{\rho}$. Our systematic analysis reveals several key trends:

\begin{itemize}
\item \textbf{Higher Effective Temperatures during the evolution:} The onset of helium rain injects heat into the envelope, temporarily boosting the planet’s effective temperature. This effect is more pronounced in models with lower $R_{\rho}$, where reduced compositional gradients allow more efficient convective transport of this heat to the surface. We find that the Schwarzschild-stable models ($R_{\rho} = 0$) must begin with lower initial entropies to ensure fuzzy core stability, especially in the presence of strong mixing in the outer envelope. As a result, despite the stronger post-rain heating, their overall effective temperatures remain lower than in Ledoux-stable models, which start with higher initial entropy. Thus, the Ledoux models still exhibit the highest effective temperatures over most of the evolution. This behavior is illustrated in the top-right panels of Figures~\ref{fig:jupiter_rho_comp} and~\ref{fig:saturn_rho_comp}. Importantly, the final effective temperatures at 4.56 Gyr for all models are consistent with current observational constraints.

\item \textbf{Enhanced Surface Metallicity:} Decreasing ${R_{\rho}}$ increases the degree of mixing between the deep interior and the outer envelope. This upward transport of heavy elements enriches the atmospheric metallicity, particularly in fully convective (Schwarzschild) models, where envelope mixing is most efficient.

\item \textbf{Lower Initial Entropy Required:} To preserve a stable dilute core within models that allow greater mixing, we found that the initial entropy distribution must be reduced. This can be seen on the bottom left panel of Figures \ref{fig:jupiter_rho_comp} and \ref{fig:saturn_rho_comp}. Higher initial entropies would erase composition gradients that help maintain the core-envelope transition, undermining the stability of the fuzzy core. In Figure \ref{fig:jup-entropies.pdf}, the initial entropies of our Jupiter models are compared to the start entropies from formation models from \cite{Cumming2018} and \cite{Muller2020}. 

\begingroup
\small
\setlength{\tabcolsep}{3pt}
\renewcommand{\arraystretch}{1.2}
\begin{table*}
\caption{Comparison of measured quantities and our final evolutionary model results at 4.56 Gyr for Jupiter and Saturn, across various $R_{\rho}$. {The brackets represent percentage deviation in units of the measurement's 1-$\sigma$ uncertainty}. References:[1] \citet{Li2012}, [2] \citet{Seidelmann2007}, [3] \citet{vonZahn1998}, [4] \citet{Guillot2023}, [5] \citet{Iess2018}, [6] \citet{Li2010}, [7] \cite{Achterberg2020,Koskinen2018}, [8] \citet{Iess2019}.}
\label{tab:jup_sat_comparison}
\centering
\begin{tabular}{@{}llcccccc@{}}
\hline
\hline
Planet & Quantity & Measurement & $R_{\rho}=0$ & $R_{\rho}=0.25$ & $R_{\rho}=0.5$ & $R_{\rho}=0.75$ & $R_{\rho}=1.0$ \\
\hline
\multirow{6}{*}{Jupiter}
& T$_{\rm eff}$ [K] & 125.57$\pm$0.07 [1] & 124.66 (0.7\%) & 125.15 (0.3\%) & 125.24 (0.2\%) & 124.54 (0.8\%) & 124.64 (0.7\%) \\
& R$_{\rm eq}$ [km] & 71492$\pm$4 [2] & 71109.96 (0.5\%) & 71579.68 (0.1\%) & 71618.98 (0.2\%) & 71877.61 (0.5\%) & 72366.23 (1.2\%) \\
& Y$_{\rm atm}$ & 0.234$\pm$0.005 [3] & 0.233 (0.2\%) & 0.238 (0.8\%) & 0.238 (0.8\%) & 0.237 (0.6\%) & 0.234 (0.2\%) \\
& Z$_{\rm atm}$ [Z$_\odot$] & $\sim 3$ [4] & 3.66 & 3.74 & 3.73 & 3.55 & 3.12 \\
& $J_2 \times 10^6$ & 14696.57 [5] & 14679.80 (0.11\%) & 14691.61 (0.03\%) & 14673.15 (0.15\%) & 14724.58 (0.19\%) & 14638.23 (0.30\%) \\
& $J_4 \times 10^6$ & –586.61 [5] & –589.37 (0.47\%) & –585.82 (0.13\%) & –586.04 (0.09\%) & –587.25 (0.11\%) & –583.41 (0.50\%) \\
\hline
\multirow{6}{*}{Saturn}
& T$_{\rm eff}$ [K] & 96.67 [6] & 96.61 (0.06\%) & 97.45 (0.8\%) & 96.75 (0.08\%) & 96.90 (0.2\%) & 97.29 (0.6\%) \\
& R$_{\rm eq}$ [km] & 60268$\pm$4 [2] & 60149.56 (0.2\%) & 60817.50 (0.9\%) & 60660.33 (0.6\%) & 60674.67 (0.2\%) & 60584.71 (0.5\%) \\
& Y$_{\rm atm}$ & 0.16–0.22 [7] & 0.168 & 0.160 & 0.177 & 0.193 & 0.204 \\
& Z$_{\rm atm}$ [Z$_\odot$] & 5–10 [7] & 8.95 & 7.15 & 7.05 & 6.34 & 6.63 \\
& $J_2 \times 10^6$ & 16290.57 [8] & 16379.42 (0.50\%) & 16305.58 (0.09\%) & 16275.89 (0.09\%) & 16283.62 (0.20\%) & 16160.87 (0.70\%) \\
& $J_4 \times 10^6$ & –935.31 [8] & –836.84 & –824.68 & –809.67 & –810.25 & –784.11 \\
\hline
\end{tabular}
\end{table*}
\endgroup

\begin{figure}
    \centering
    \includegraphics[width=1.04\linewidth]{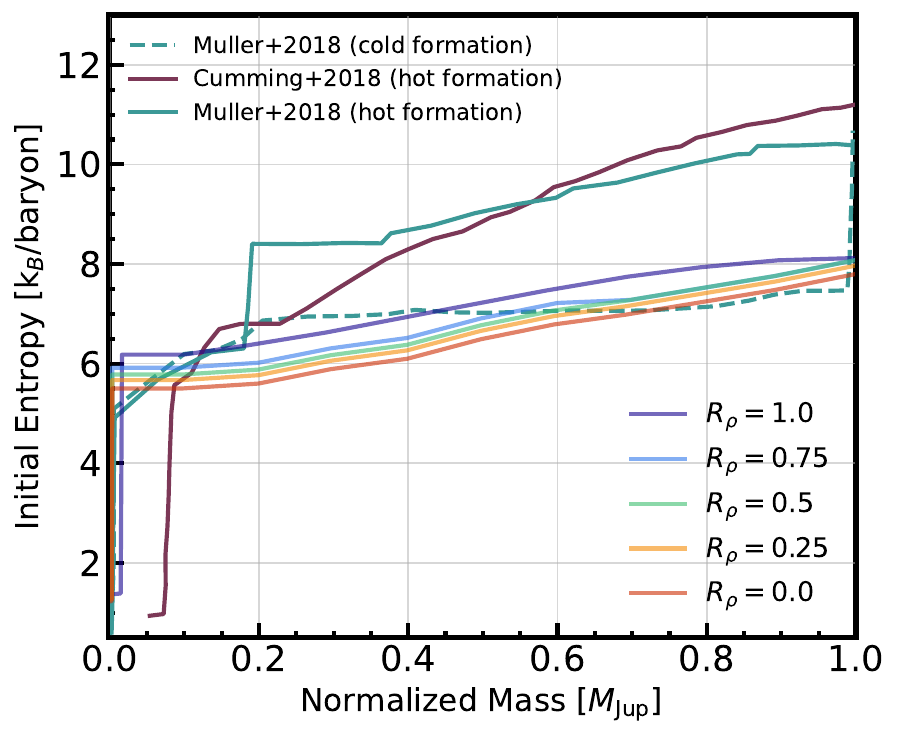}
    \caption{Initial entropy profiles for our Jupiter models with varying $R_{\rho}$ values, compared against hot-start (solid lines) and cold-start (dashed lines) initial conditions from \citet{Cumming2018} (maroon) and \citet{Muller2020} (green). Our results show that models initialized with cold-start conditions yield the best match to Jupiter’s present-day observables.}
    \label{fig:jup-entropies.pdf}
\end{figure}

\item \textbf{Cooler Interior Temperatures:} Although the surface entropy converges to $\sim6.2$ k$_\mathrm{B}$/baryon for Jupiter across all models by 4.56 Gyr, the internal temperature structure varies. Models with lower ${R_{\rho}}$ exhibit cooler interiors due to more efficient convective heat transport, which flattens temperature gradients and facilitates thermal equilibration throughout the planet. For Saturn, we also notice that the models undergo compressional heating at the center as the planet cools and decreases in size.

\item \textbf{Extent of Fuzzy Core and Reduced Heavy-Element Mass Requirement:} The total heavy-element mass needed to satisfy Jupiter’s observational constraints decreases from 44.25\,\mearth\ in Ledoux-stable models to 38\,\mearth\ in Schwarzschild-stable models. For Saturn, the total mass varied from 25.5 to 24.5 \mearth. This trend reflects the more extensive redistribution of heavy elements throughout the envelope at low ${R_{\rho}}$, which raises the envelope’s metallicity without requiring additional heavy-element mass. Models with excessive mixing tend to exceed the observational constraint of $\sim 5\,Z_\odot$ for Jupiter. Additionally, Schwarzschild models for both planets require a smaller compact core mass compared to their Ledoux counterparts. Despite these changes, the radial extent of the fuzzy core remains nearly identical across all models, as it is constrained by the need to reproduce the measured gravitational moments.

\item  \textbf{Closer Match to \brunt frequency for Saturn}: The \brunt frequency ratio profiles for the various Jupiter and Saturn models are shown in Figure~\ref{fig:brunt}. The Brunt in Saturn is substantial and extends to more than 60\% of its radius, aligning well with the post-Cassini ring seismology analysis of \citet{Mankovich2021}. As $R_{\rho}$ decreases, two distinct regions of enhanced Brunt frequencies emerge: one within the dilute core and another in the helium rain layer. The mean Brunt frequency ratio increases as we move from $R_{\rho} = 1$ (Ledoux) to $R_{\rho} = 0$ (Schwarzschild), indicating stronger stable stratification in key interior regions.
\end{itemize}

\begin{figure*}%
\centering
  \subfigure[Jupiter]{%
    \includegraphics[width=0.48\textwidth]{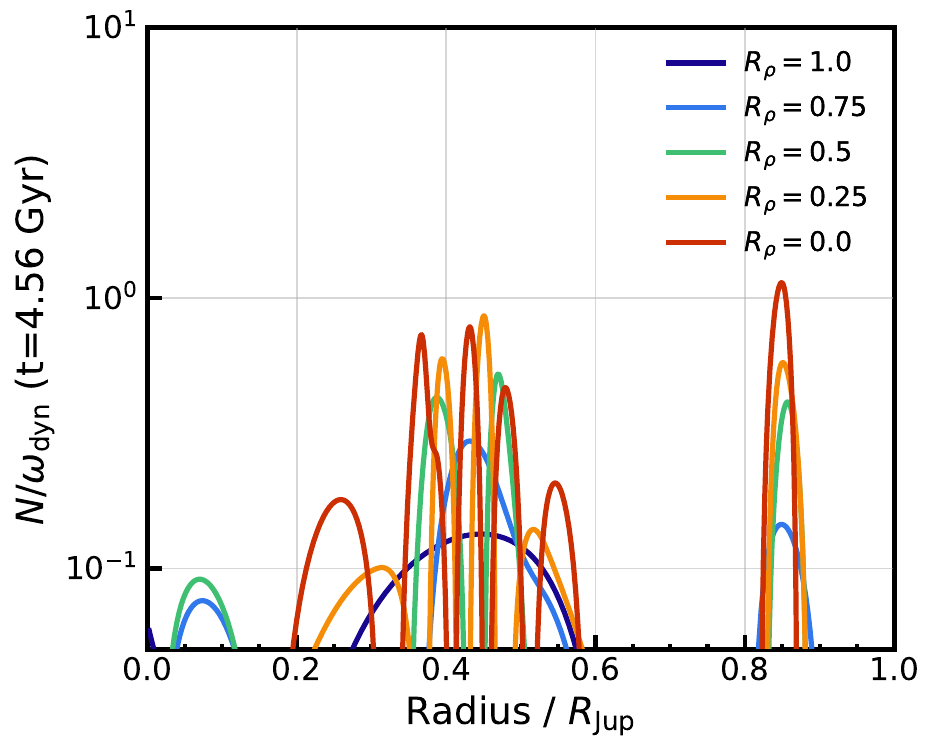}
    \label{fig:jup_brunt}
  }\hspace{0.01\textwidth}
  \subfigure[Saturn]{%
    \includegraphics[width=0.48\textwidth]{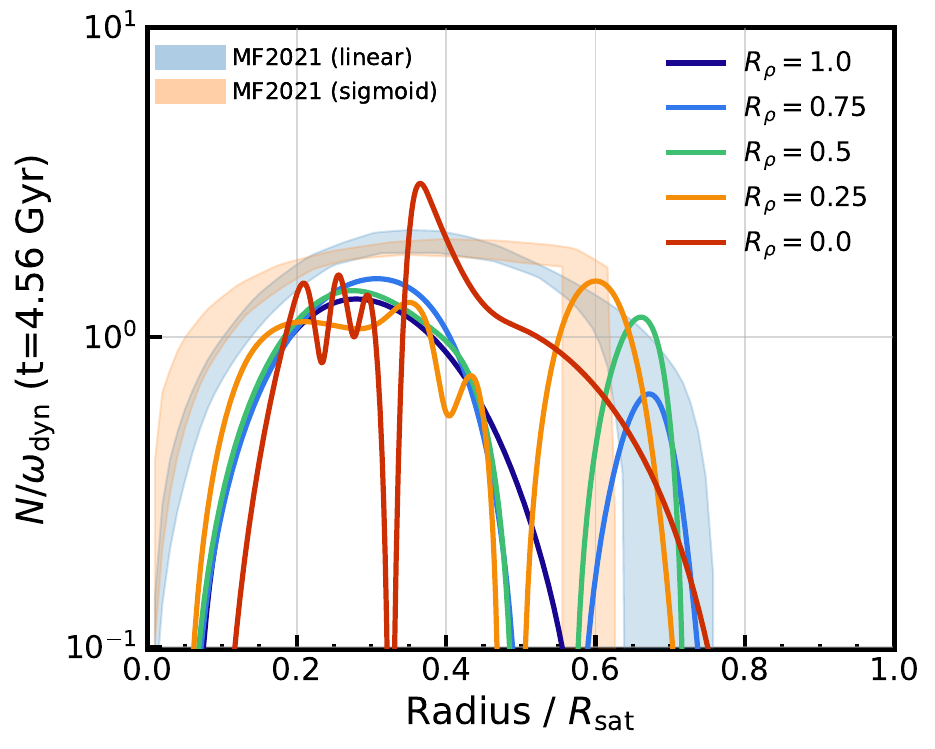}
    \label{fig:sat_brunt}}
    \caption{\brunt frequency ratios for Jupiter and Saturn at 4.56 Gyr for models with varying $R_{\rho}$. The shaded region in panel b indicates the seismologically inferred stratified zone from \citet[MF2021]{Mankovich2021}, with a characteristic ratio of 2 extending to 60\% of Saturn’s radius. Our models reveal comparable stable regions—one associated with the dilute core and another arising from the helium rain layer in both planets. The location of the peaks of our distributions for Saturn matches well with the predictions of MF2021. The raw brunt data have been smoothed out for better visualization in this plot.}%
    \label{fig:brunt}%
\end{figure*}

\section{Discussions and Conclusions}
\label{sec:conclusions}

In this paper, we have generalized our earlier work on the thermal and structural evolution of Jupiter and Saturn, which assumed only Ledoux ($R_{\rho} = 1$) convection, to explore the dependence upon the parameter $R_{\rho}$. This has allowed a study comparing a range of convective thermal and compositional transport efficiencies spanning the range between the inefficient Ledoux and {efficient} Schwarzschild ($R_{\rho} = 0$) limits. Using the latest hydrogen–helium equation of state, updated atmospheric boundary conditions, and the \texttt{APPLE} evolution code \citep{Sur2024_apple}, we constructed evolutionary models that self-consistently include helium rain and preserve a fuzzy heavy-element core over 4.56 Gyr. Crucially, we move beyond Ledoux-stable models by systematically varying $R_{\rho}$ from 1 (Ledoux criterion) to 0 (Schwarzschild criterion), capturing a range of convective mixing regimes and identifying their impact on observable quantities and the required initial configurations. In contrast to our earlier paper, this study directly explores the implications of using not only the Schwarzschild criterion for convective stability, but also other non-unity values of $R_{\rho}$. This allows one to estimate the potential range of evolutionary behaviors and initial model profiles allowed if semiconvection (not yet easily incorporated into evolutionary codes with sufficient numerical fidelity) operates in such planets. We show that though such approximate models remain viable, they motivate further study into the physics and implementation of semiconvection and convective overshoot in giant planet evolutionary calculations \citep{Chabrier2007, Leconte2012, Garaud2017, Tulekeyev2024}. 

{In the absence of a fully self-consistent treatment of semiconvection in one-dimensional planetary evolution codes, we have built upon the parameterized framework adopted in previous studies of Jupiter and Saturn’s thermal evolution \citep{Mankovich2016, Mankovich2020, Howard2024, Sur2024_apple, Nettelmann2025}. In this approach, the global parameter $R_{\rho}$, which is different from the density ratio $R_0$, is used to bracket the two limiting regimes of convective stability: the Schwarzschild criterion (efficient convection) and the Ledoux criterion (stability due to compositional gradients). We have extended this prescription by systematically varying $R_{\rho}$ to explore a broad range of mixing efficiencies and thermal states in models with fuzzy cores—an aspect not previously addressed. Nonetheless, we acknowledge that a fully self-consistent theory of semiconvection remains an outstanding challenge in planetary modeling, and our simplified approach should be viewed as an interim step toward that goal.}

In broad summary, we find that the onset of helium rain enhances the planet's effective temperature, but the overall thermal evolution of the planet is governed mostly by the initial entropy structure, which must be uniformly lower in Schwarzschild-stable models to preserve dilute cores. As a result, Ledoux-stable models tend to exhibit higher internal temperatures throughout evolution. Nevertheless, by construction, the final effective temperatures at 4.56 Gyr remain consistent with observations for both planets across all $R_{\rho}$. The Schwarzschild models yield reduced total heavy-element masses and smaller compact cores, while preserving the observed radii and gravitational moments. Surface metallicities evolve in tandem with mixing efficiency: stronger convective mixing enriches the outer envelope while still matching the Galileo \citep{vonZahn1992, vonZahn1998} and Cassini \citep{Spilker2019, Iess2018, Iess2019} constraints.  In Saturn, we find that helium rain is halted at the edge of the dilute core, causing helium to accumulate in an intermediate region. If the dilute core is fully eroded by convection over time \citep{Tulekeyev2024}, a helium-rich layer—or ``helium ocean"—may form in the deep interior \citep{Howard2024}.

A particularly promising diagnostic is the \brunt frequency profile. For Saturn, our $R_{\rho} \ne 1$ models show two distinct regions of stable stratification--one associated with the fuzzy core and the other with the helium rain layer, especially in our Schwarzschild model. We note that the magnitude and shape of the Brunt frequency profiles depend sensitively on the assumed initial distributions of entropy and heavy elements. \footnote{In this spirit, we note that our current best-fit $R_{\rho}=1$ model differs slightly from that in \citet{Sur2025} due to the different function form of the initial entropy profiles employed in the current study.} The resulting Brunt frequency ratios align closely with those inferred from Cassini ring seismology \citep{Mankovich2021} extending up to 60\% of its radius, and growing stronger as $R_{\rho}$ decreases. This agreement provides a constraint on the interior stratification of Saturn and further supports the presence of a layered structure shaped by composition gradients. Similar detections of oscillation modes in Jupiter \citep{Xu2025} would offer a valuable window into its internal structure and further test the predictions of our models. 

We note that \citet{Anders2022} demonstrated in idealized 3D simulations that convective overshoot can erode semiconvective layers over time, leading to convergence between the Ledoux and Schwarzschild criteria. However, their study focused on a narrow parameter space relevant to stellar interiors, using the Boussinesq approximation and neglecting key planetary processes such as compressibility, rotation, and helium phase separation. While suggestive, such results must be applied to gas giants with caution. Though not definitive, our models offer a pathway to connect these dynamical processes with observable features, particularly when interpreted in light of Cassini’s gravity and seismology constraints. 

In summary, we have presented a unified framework for understanding the thermal evolution and compositionally stratified interiors of Jupiter and Saturn across a range of convective algorithmic parameterizations. While many questions remain --- particularly regarding compositional evolution and phase separation --- we suggest that our findings represent yet another useful step toward unraveling the internal histories of the solar system’s giant planets.

\section{Acknowledgment}   
Funding for this research was provided by the Center for Matter at Atomic Pressures (CMAP), a National Science Foundation (NSF) Physics Frontier Center, under Award PHY-2020249. Any opinions, findings, conclusions, or recommendations expressed in this material are those of the author(s) and do not necessarily reflect those of the National Science Foundation. 
YS is supported by a Lyman Spitzer, Jr. Postdoctoral Fellowship at Princeton University.

\bibliography{references}{}
\bibliographystyle{aasjournal}

\end{document}